\documentclass[a4paper,11pt]{article}
\usepackage{pos}

\title{Charged lepton flavour violation in muons: theory overview}

\author*{Ana M. Teixeira}

\affiliation{Laboratoire de Physique de Clermont (UMR 6533), CNRS/IN2P3,\\
Univ. Clermont Auvergne, 4 Av. Blaise Pascal, 63178 Aubi\`ere Cedex,
France
}

\emailAdd{ana.teixeira@clermont.in2p3.fr}

\abstract{We briefly overview some theoretical aspects of charged lepton flavour violation in rare muon transitions and decays. Relying on the effective field theory approach (model-independent), we discuss the probing power of charged lepton flavour violation processes regarding the new physics scale, especially in what concerns muon-electron conversion in nuclei. We then illustrate how these processes can provide valuable information (allowing to constrain, falsify - or conversely to learn about the underlying new physics properties) for a few examples of well-motivated extensions of the Standard Model. We finally comment on how leptonic CP violation should be carefully considered upon studies of charged lepton flavour violation in extensions of the Standard Model featuring heavy sterile fermions.}

\FullConference{Muon4Future Workshop (Muon4Future2023)\\
 29–31 May, 2023\\
Venezia, Italy\\}

\begin{document}
\maketitle

\section{Introduction}\label{sec:intro}
Flavour physics has played a key role in the shaping and evolution of particle physics since the early days of the Standard Model (SM). In recent years, flavours have paved the way to searches for New Physics (NP): from the discovery of neutrino oscillations, to the long-standing tensions between theory and observation regarding the anomalous magnetic moment of the muon, to conflicting data on possible violation of lepton flavour universality in meson decays. Curiously, many of these tensions are deeply rooted in the lepton sector. 
Strictly forbidden in the context of the SM, the violation of lepton flavour was clearly demonstrated with neutrino oscillation phenomena, and in principle, once the latter occurs, nothing precludes the violation of charged lepton flavours. However, minimal extensions of the SM envisaged to accommodate neutrino oscillation data (via the addition of massive Dirac neutrinos, and lepton mixings) lead to tiny rates for charged lepton flavour violating (cLFV) transitions. Thus, any observation of a cLFV process would offer striking evidence for NP - beyond such minimal SM extensions. 

In view of their very appealing properties, muons are ideal objects to search for cLFV, with numerous muonic transitions as the object of extensive dedicated searches at low-energies and at colliders. The former include radiative decays ($\mu \to e \gamma$), three-body decays ($\mu \to 3 e$), muonic-bound state processes, such as Muonium (Mu $=\mu^+ e^-$) oscillations and decays, or transitions in the presence of muonic atoms (including neutrinoless $\mu-e$ conversion, among others). 
At higher energies - at the LHC or at future lepton colliders -, numerous processes can also lead to a discovery of cLFV, as is the case of neutral boson decays ($Z, H \to \mu e$), or the study of high-$p_T$ dilepton tails in $pp \to \mu \ell$. 
The expected future sensitivities of muon-cLFV dedicated facilities suggest that these processes might offer a sensitivity to NP mediators up to scales around $10^5$~TeV~\cite{EuropeanStrategyforParticlePhysicsPreparatoryGroup:2019qin}. 

Muon cLFV is thus poised to offer evidence for NP, or then further strengthen the constraints on the scale of the new mediators and on the intensity of the new interactions with leptons; from a theory perspective, studies of muon cLFV can be carried out following an effective (model-independent) approach, or investigating the prospects for cLFV of a given, well-motivated NP model (especially those which also offer an explanation to the SM observational issues, including neutrino mass generation and the baryon asymmetry of the Universe, or a viable dark matter candidate). In what follows, we will illustrate these approaches.

\section{Muon cLFV: effective approach to NP}\label{sec:muonEFT}
\begin{figure}[h!]
    \centering
    \mbox{
    \includegraphics[width=0.40\textwidth]{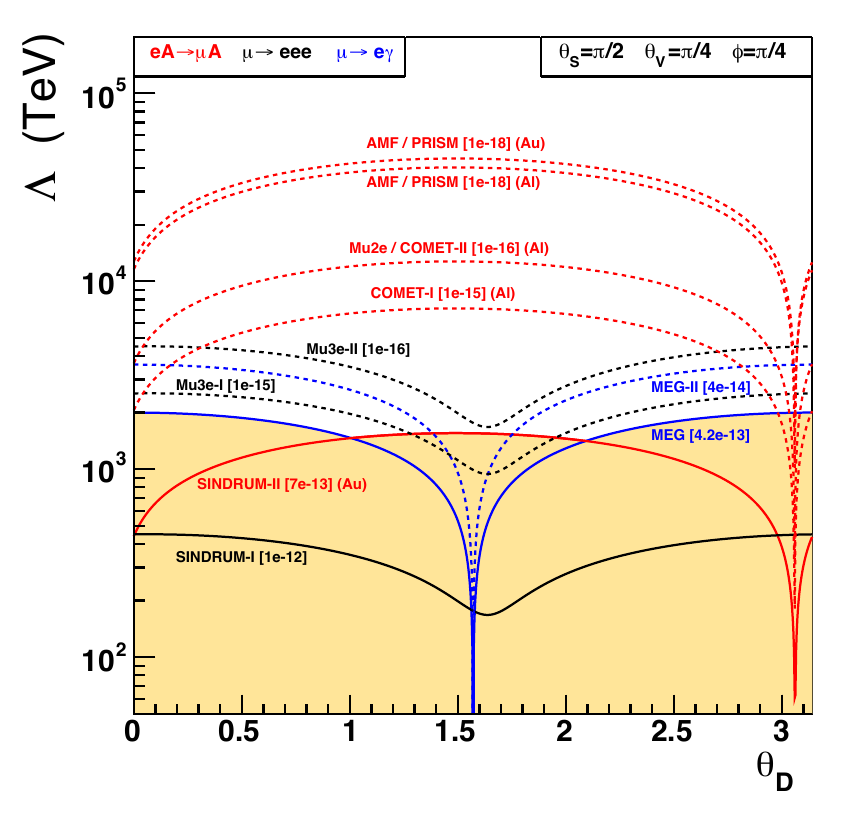}\hspace*{8mm} 
    \includegraphics[width=0.52\textwidth]{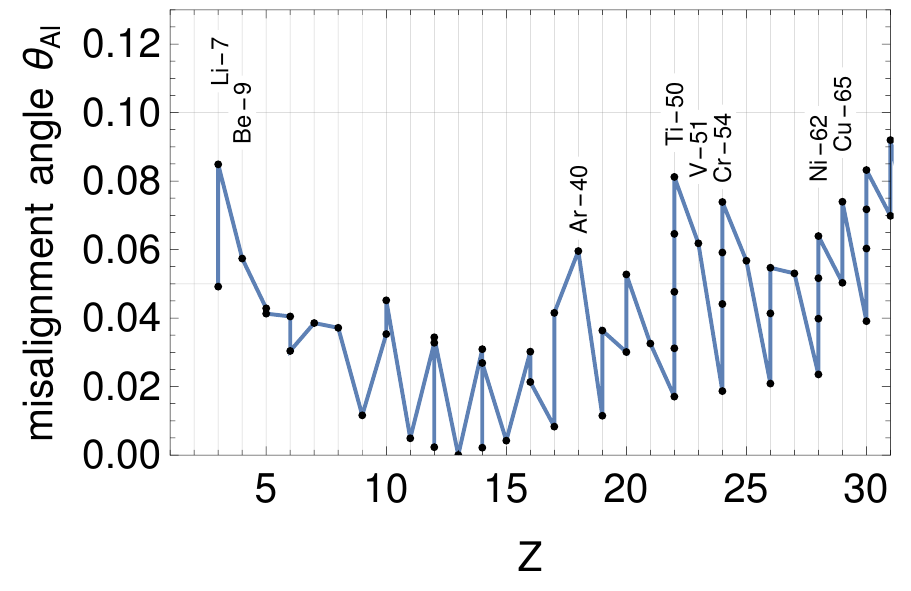}}
    \caption{On the left, sensitivity to the NP scale of different cLFV muon transitions, as a function of the ``dipole'' dominance parameter $\theta_D$: radiative decays (blue), three-body decays (black) and neutrinoless conversion in nuclei (red); solid (dotted) lines denote current bounds (future experimental sensitivity); from~\cite{Davidson:2022nnl}. 
    On the right, complementarity of the information for distinct nuclei ($Z$) regarding the NP at the source of $\mu-e$ conversion, with respect to that of Aluminium targets; from~\cite{Heeck:2022wer}.    
    }
    \label{fig:eft-scales-targets}
\end{figure}

In the effective field theory (EFT) approach, the SM Lagrangian is generalised to include high-order, non-renormalisable terms that encode the effects of heavy new states (and their interactions), possibly present at scales far above the electroweak one, $\Lambda_\text{NP} \gg \Lambda_\text{EW}$. The observables can be expressed in terms of the new couplings and scale, and available bounds are then used to constrain combinations of the latter. In view of the extensive number of operators that can be at the origin of the cLFV transitions (dimension 6, and higher), the task is far from simple. In recent years, numerous analyses have emphasised the relevance of taking into account higher-order contributions, renormalisation group running (leading to operator mixing), among other points, also suggesting the importance of including as many observables and operators as possible, so to more efficiently constrain the nature of the NP contributions (see, e.g.~\cite{Crivellin:2017rmk}). 
A new parametrisation proposal~\cite{Davidson:2022nnl}, relying on projections onto a space spanned by the most relevant operators for muon cLFV transitions, $\vec C = \{C_D, C_S, C_{VR}, C_{VL}, C_\text{N-light}, C_\text{N-heavy} \}$, allowed to infer the sensitivity to NP scales of data (current bounds and future sensitivities) stemming from searches for radiative and 3-body decays, and conversion in nuclei. As can be seen from the left panel of Fig.~\ref{fig:eft-scales-targets}, current bounds from MEG on $\mu \to e \gamma$ and from SINDRUM on neutrinoless conversion in Gold nuclei allow to draw limits on the scale of cLFV new physics up to $\mathcal{O}(10^3~\text{TeV})$, respectively for the cases of dipole or 4-fermion dominated NP contributions ($\theta_D \to 0, \pi$ or $\theta_D \to \pi/2$). In the future, and for both regimes, the expected sensitivity of searches for $\mu-e$ conversion in Aluminium by both COMET and Mu2e should allow probing scales $\Lambda_\text{NP} \sim 10^4$~TeV (or even $10^5$~TeV for an advanced muon facility)~\cite{Davidson:2022nnl}.

The impressive sensitivity associated with $\mu-e$ conversion searches lends hope to a future observation of these rare transitions. In the event of a discovery, it becomes important to (re)visit the choice of nuclear target, so as to infer as much information as possible on the nature of the underlying NP. Recent studies suggest that Vanadium-51 or Lithium-7 are feasible choices of targets, which could offer the most complementary (or orthogonal) information to that of Aluminium targets~\cite{Heeck:2022wer}. The potential complementarity of several  targets is displayed in the right panel of Fig.~\ref{fig:eft-scales-targets}.

A very interesting process, also occurring in muonic atoms, is that of the flavour and lepton number violation transition (LNV), 
$\mu^- + (A,Z) \to e^+ + (A, Z+2)^{(*)}$, in which the final state nucleus might be in an excited state.
This transition could  offer a pivotal connection between cLFV and LNV (by two units), in association with the possible Majorana nature of the NP mediator, and a link to the mechanism of neutrino mass generation). Already experimentally searched for, this process is highly non-trivial to tackle from a theory point of view: not only does it call upon higher-order operators (contrary to the ``usual'' cLFV dimension 6 operators) but, as of today, the nuclear matrix elements remain extremely challenging to compute, rendering any inference hard to draw.

\section{Muon cLFV - learning about models of NP}\label{sec:muonNPmodels}
If the SM extension includes sources of cLFV, the model can in principle account for extensive ranges for the cLFV observables. Depending on the dominant contributions to given transitions, one can nevertheless try to identify distinctive patterns for the observables: for example, should one have photon-penguin dominance in $\mu-e$ conversion, then one can expect a correlation between the former and radiative (as well as 3-body) decays; should there be a $Z$ penguin dominance, then it is likely that BR($\mu \to  3 e$) and CR($\mu-e$, N) be correlated. The confrontation of ratios of future measurements of observables with 
the predictions of a given model might offer valuable information to falsify a given construction, or to strengthen its viability. Several examples have been presented in~\cite{Calibbi:2017uvl}. 
Here we very briefly consider a few (recent) examples of how cLFV - especially through the muon channels - can offer powerful probes of NP models. We consider illustrative cases associated with mechanisms of neutrino mass generation, grand unified theories (GUT), or those aiming at explaining current tensions of the SM.

\begin{figure}[h!]
    \centering
    \mbox{
    \includegraphics[width=0.46\textwidth]{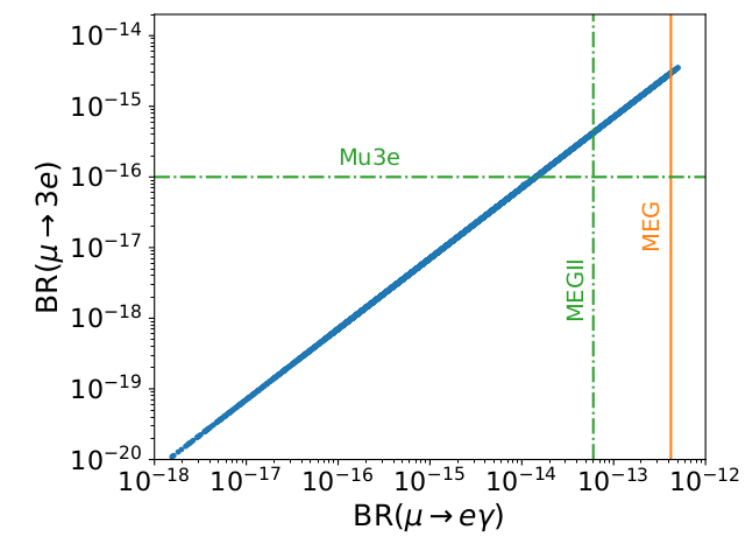}\hspace*{8mm} 
    \includegraphics[width=0.48\textwidth]{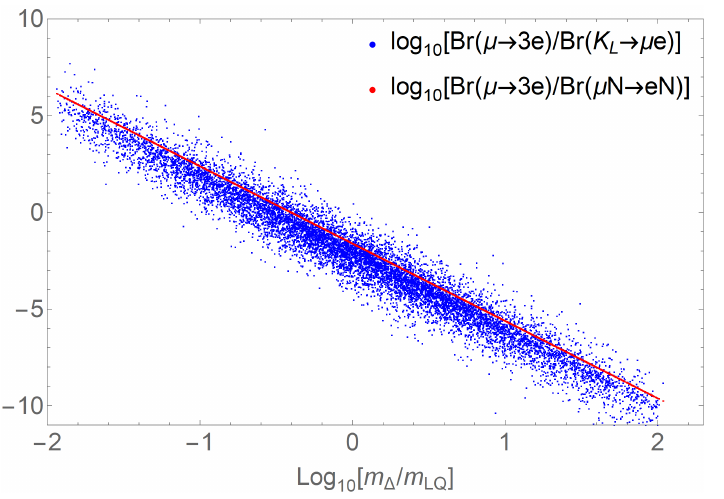}}
    \caption{On the left, correlated behaviour of $\mu \to e \gamma$ and $\mu \to 3 e$ decays for a scotogenic NP model, as studied in~\cite{Alvarez:2023dzz}. 
   On the right, ratios of cLFV observables (BR($\mu \to 3e$), BR($K_L \to \mu e$) and CR($\mu-e$, N)) as a function of the ratio of the NP mediator masses (scalar triplet and LQ masses) in an SU(5) GUT model; from~\cite{Calibbi:2022wko}.
    }
    \label{fig:models:scoto-gut}
\end{figure}

Scotogenic models offer a powerful link between two of the SM's most pressing issues - massive neutrinos and the need for a dark matter (DM) candidates. Furthermore, scotogenic models usually call upon symmetries to stabilise the DM candidate, which in general prevents tree-level neutrino mass generation (which then occurs at 1-loop or at higher levels). \\
\noindent
Many scotogenic models lead to interesting cLFV signatures: a recent analysis~\cite{Alvarez:2023dzz} considered an extension of the SM via 2 Weyl fermions as well as Majorana fermions (singlets and scalars); such a construction was shown to successfully explain neutrino oscillation data, put forward a viable DM candidate, and further saturate the tensions on the muon anomalous magnetic moment. In the muon-electron sector, cLFV 3-body decays are dominated by photon-penguin contributions, leading to a strict correlation between BR($\mu \to e\gamma$) and BR($\mu \to 3e$), as shown in the left panel of Fig.~\ref{fig:models:scoto-gut}; a future determination of these observables might allow to falsify such a NP scenario. 

Another enlightening example can be found in GUT models, ambitious constructions which aim at addressing not only the SM observational issues, but also at overcoming some of its theoretical caveats. Despite their appeal, many such models rely on extended field contents (which can be present at very high scales), rendering them hard to probe. However, certain GUT models lead to interesting cLFV ``signatures'', which offer a means to learn about the new spectrum (among other features). This was explored in~\cite{Calibbi:2022wko}, in which a type II seesaw was embedded into an SU(5) non-supersymmetric GUT. Dedicated cLFV studies revealed peculiar patterns, dictated by the mass ratio of the new triplet seesaw ($\Delta$) mediator and its SU(5) leptoquark (LQ) partners. As visible on the right panel of Fig.~\ref{fig:models:scoto-gut}, there is a clear correlation of BR($\mu \to 3e$), BR($K_L \to \mu e$) and CR($\mu-e$, N). A near-future observation (as well as current bounds) seem to prefer scenarios in which $M_\Delta > M_\text{LQ}$. Furthermore, cLFV observations can also hint on the ordering of the light neutrino spectrum: if  BR($K_L \to \mu e$)~$> 10 \,\times\, $~CR($\mu-e$, N), the inverted ordering would be disfavoured. 

The previous example also highlights how other cLFV transitions (in addition to the well-known muon radiative and 3-body decays, and conversion in nuclei) might be instrumental in probing NP models. These modes include (semi-)leptonic meson decays, as well as the already mentioned Muonium cLFV decays. Likewise, the tau sector should also be included, even the more involved decay modes (e.g. doubly lepton-flavour violating $\tau \to \mu \bar e \mu$ 3-body decays), which can sometimes be at the source of the most stringent constraints (see, for instance, the study of~\cite{Kriewald:2022erk}).

\section{Leptonic phases and muon cLFV }\label{sec:CPVmuoncLFV}
As mentioned before, ``patterns'' of cLFV observables have been intensively explored as means to falsify numerous NP models - including mechanisms of mass generation. In general, the different fundamental seesaw realisations lead to well-defined signatures of cLFV ratios, which reflect the nature of the distinct mediators, and whether the transitions occur at tree-level or at higher orders. Especially in the case of low-scale seesaw realisations, the contributions to the distinct cLFV transitions can be sizeable, and thus lead to important constraints on these SM extensions. 

\begin{figure}[h!]
    \centering
    \mbox{
    \includegraphics[width=0.48\textwidth]{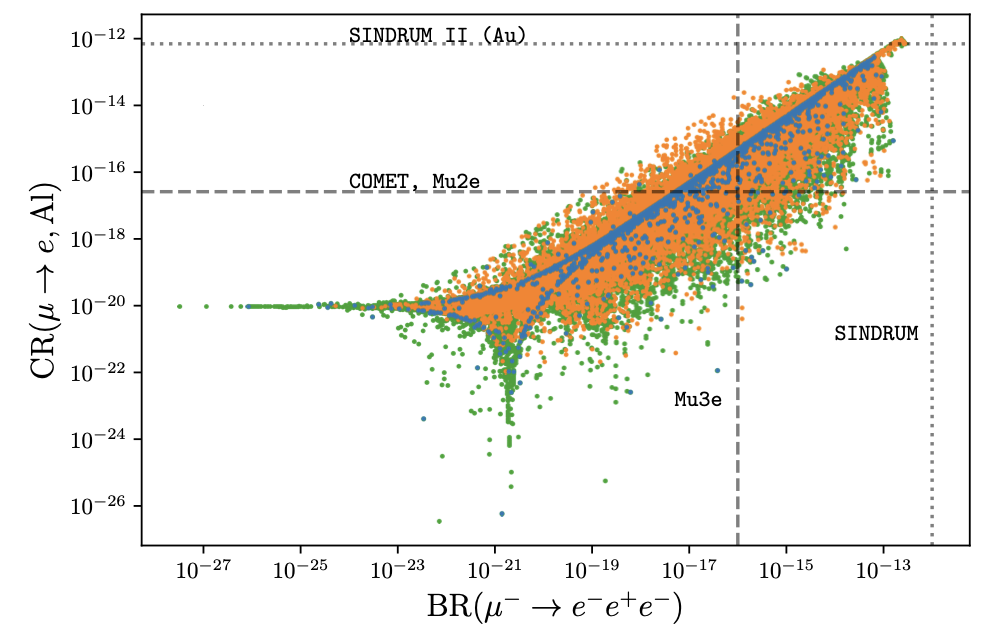}\hspace*{5mm} 
    \includegraphics[width=0.47\textwidth]{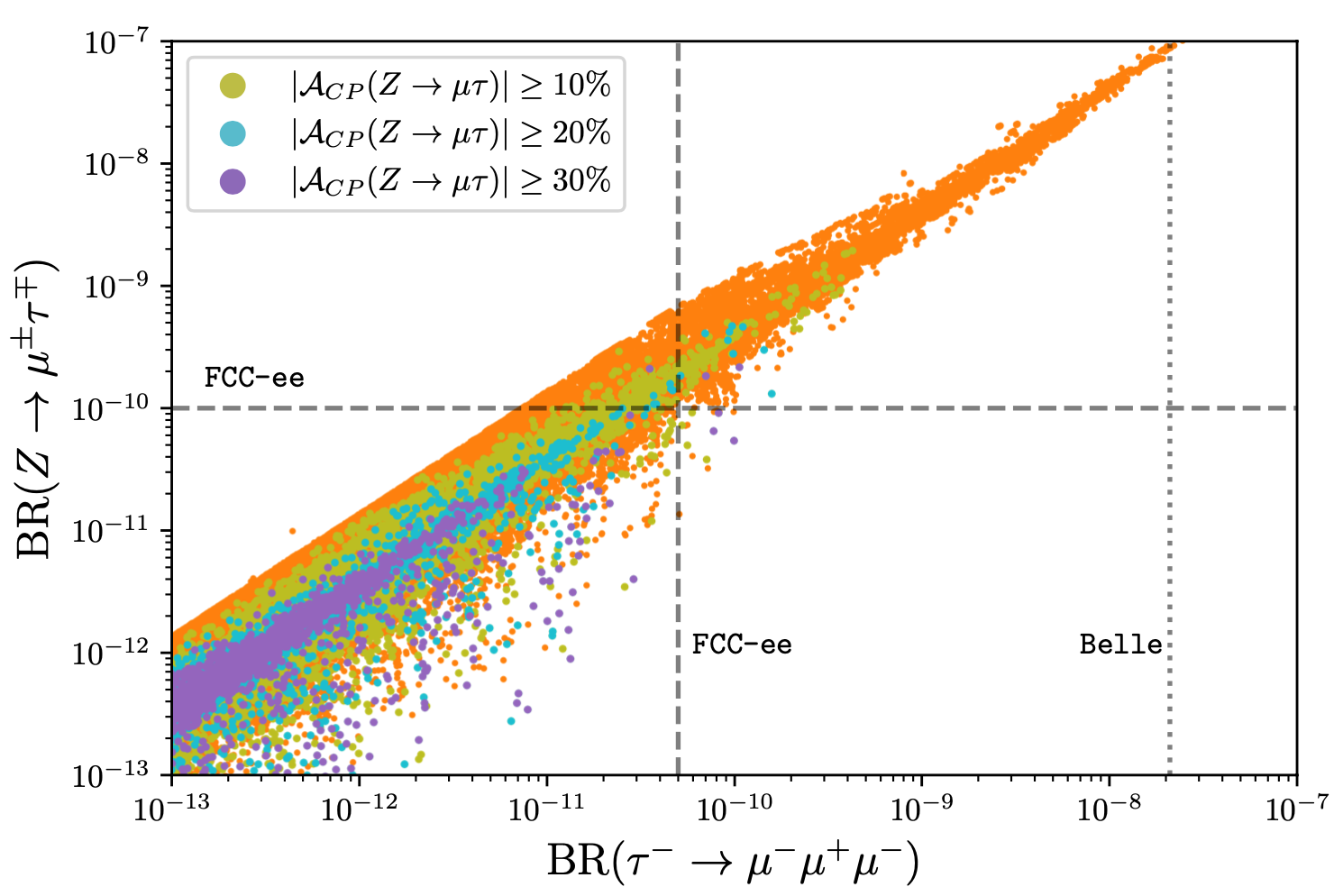}}
    \caption{On the left, projections for CR($\mu-e$, Al) vs. BR($\mu \to 3e$), for a minimal (ad-hoc) SM extension via 2 HNL: CP conserving limit (blue) and the effect of CP violating leptonic phases (orange and green); dotted (dashed) lines denote current bounds (future sensitivity); from~\cite{Abada:2021zcm}. 
    On the right, prospects for BR($\tau \to 3\mu$) vs. BR($Z\to \tau\mu$), also for a SM extension via 2 HNL. The colour code denotes the CP-asymmetry in $Z\to \tau\mu$ decays; dotted (dashed) lines denote current Belle bounds (FCC-ee future sensitivity); from~\cite{Abada:2022asx}.   
    }
    \label{fig:CPV-cLFV}
\end{figure}

However, most studies rely on CP conserving interactions - but leptonic CP phases (be them Dirac and/or Majorana) are in general present in SM extensions aiming at accounting for neutrino masses and lepton mixings, and are a crucial ingredient to explain the baryon asymmetry of the Universe via leptogenesis. Recently, the impact of the leptonic CP phases on cLFV observables was addressed for the generic case of SM extensions via at least 2 Majorana heavy neutral leptons (HNL), relying on a simplified-model approach to well-motivated NP models, as for instance the type I seesaw and its variants~\cite{Abada:2021zcm,Abada:2022asx}. 
In association with the 2 heavy sterile states, the effects due to the new CPV phases can be at the source of important constructive and destructive interferences, which strongly alter the theoretical expectations, and the interpretation of any future data. 
This is clearly manifest in the left panel of Fig.~\ref{fig:CPV-cLFV}: for the CP conserving case (blue) there is a visible correlation of the neutrinoless conversion and 3-body muon decays (a consequence of having both processes dominated by $Z$-penguin contributions); in particular, and should HNL be at the source of cLFV phenomena, 
any future observation of $\mu \to 3 e$ would suggest that  $\mu-e$ conversion should also be observed. Once CP violation is present (green and orange), there is a dramatic loss of correlation. In view of the potential presence of CPV phases, constraining HNL extensions from cLFV observables is an exercise that must be carefully done\footnote{Although in general cLFV observables remain privileged probes for low-scale seesaw realisations relying on additional heavy sterile fermions, lepton flavour universality violation observables and $\Gamma(Z \to \text{inv.})$  can be powerful complementary probes to cLFV in the $\mu-e$ sector, as recently pointed out in~\cite{Abada:2023raf}.}. 
Leptonic CPV phases are also expected to impact cLFV (neutral) boson decays; at a future FCC-ee, one can look for their presence by considering CP asymmetries in cLFV $Z$ boson decays. In particular, and as seen from the right panel of Fig.~\ref{fig:CPV-cLFV}, for both BR($Z \to \mu \tau$) and BR($\tau \to 3 \mu$) within experimental reach, $\mathcal{A}_{CP} (Z \to \mu \tau)$ can be as large as 20\%, thus offering a potential 3-fold way to test such minimal SM extensions via at least 2 heavy sterile states.

\section{Overview}\label{sec:overview}
In view of the very ambitious experimental programme, lepton physics is expected to provide valuable hints on the underlying New Physics model, which is clearly needed to address the SM caveats. Charged lepton flavour violating muon decays and transitions will offer precious help in identifying the NP model at work, hinting on the scale and nature of the new mediators, reducing ambiguities, constraining parameter spaces,  or even falsifying SM extensions.

The near future will witness a unique experimental progress in muon-cLFV dedicated searches~\cite{M4F}, rendering a possible  a discovery of New Physics, potentially before the observation of new resonance at colliders. Theory efforts and improved theoretical approaches must keep on par with experimental developments: one must thus consider as many observables as possible (including proposals of new ones), explore distinct approaches (effective theories, and dedicated phenomenological studies of well-motivated NP models), striving to  increase theoretical control at all levels, as well as fully exploring 
synergies of observables and flavour sectors.

\acknowledgments
We acknowledge support from the EU's Horizon 2020 research and innovation programme under the Marie Sk\l{}odowska-Curie grant agreement No.~860881 (HIDDeN network) and from the IN2P3 (CNRS) Master Project, ``Flavour probes: lepton sector and beyond'' (16-PH-169).

\end{document}